# AERODYNAMIC MODELS FOR HURRICANES

## II. Model of the upper hurricane layer

A.I. Leonov

The University of Akron, Akron, OH 44325-0301

E-mail: **leonov@uakron.edu**

**Abstract**

This second paper of the series (see the first one in [1]) models the dynamics and structure of upper hurricane layer in adiabatic approximation. Formulation of simplified aerodynamic model allows analytically express the radial distributions of pressure and wind speed components. The vertical evolution of these distributions and hurricane structure in the layer are described by a coupled set of equations for the vertical mass flux and vertical momentum balance, averaged over the eye wall cross section. Several realistic predictions of the model are demonstrated, including the change of directions for the component of radial wind speed and angular velocity of hurricane with altitude.

## 2.1. Introduction

This paper analyzes the finite fluctuation of homogeneous atmosphere in the adiabatic layer caused by the hurricane rotation and buoyancy effect. The ultimate goal of this paper is establishing the space distribution of velocity components and pressure, as well as the vertical evolution of eye wall (EW) structure and flow/thermodynamic variables in the upper layer. A steady model, describing the vertical and radial distribution of dynamic variables and hurricane structure in adiabatic layer, is formulated as a simplified version of basic set of aerodynamic equations (1.7)-(1.10) from [1]. This approach is also convenient for analyzing quasi-stationary cases when the hurricane parameters are slowly changed in time.

According to Figure 2.1 the upper hurricane layer is located above the boundary layer, i.e. $z \geq h_b$. In radial direction, it consists of three regions: the eye region $0 < r < r_i$, "EW jet" region $r_i < r < r_e(z)$, and external region $r_e(z) < r < r_a$. Here $r_i$ and $r_e(z)$ are the internal and external EW jet radii, respectively, $r_a \approx const$ is the external radius of hurricane, and the height of boundary layer $h_b$ can be radius dependent in the region $r > r_e$. The common condition $r_i = const$ describes a frontal zone at this interface. We denote as $m$ and $T$ the moisture content and temperature in the EW jet, respectively. The values of these parameters are assumed to be different from the ambient values $m_a$ and $T_a$ so that $m > m_a$ and $T > T_a$. One can find in the text [2] the basic relations between thermodynamic variables (see also a brief explanation in [1]).

Underlying idea of this paper is that the large scale fields and structures in hurricane are deterministic and coherent, in spite of small scale turbulent fluctuations, confined in relatively small zones. Therefore this paper does not formally employ turbulent approaches, commonly utilized in the majority of papers modeling hurricanes (e.g. see [3-5]), although it sometimes uses turbulence concept to explain specific effects. So the mass and heat transfer between the EW jet and exterior/interior rotating air is neglected. Yet, we widely employ the concept of Kelvin-Helmholtz stability [6], which demands continuous distributions of velocity and pressure fields in flows of ideal fluids. The thin zones, where continuity is impossible, are expected to create turbulization.

The simple aerodynamic approach, used in the paper for analyzing steady axisymmetric flow of ideal gas in hurricane, also has an advantageous feature – existence of a first integral, which express angular momentum as an arbitrary function of the stream function (see formula (1.12) in [1]).

The paper is organized as follows. The Section 2.2 describes the simplifications of aerodynamic equations used for formulation of the model. Section 2.3 presents formulas for radial distribution of basic variables. Section 2.4 derives and analyzes the equation describing the vertical evolution of EW jet. Section 2.5 estimates model parameters and discuss results of numerical illustrations. Conclusive remarks are given in Section 6.

## 2.2. Model formulation

Adiabatic approximation is the most serious assumption made in the present modeling. It neglects the possible additional condensation process in the upper layer, as compared with the major condensation just below this layer. The model below demonstrates the validity of aerodynamic approach using some simplified equations, which make possible almost analytical study. Along with adiabatic, two other modeling assumptions relevant to observations are:

1. The vertical velocity component $u_z$ is confined only in the EW jet region $r_i < r < r_e$ and is negligible in the inner and outer regions of hurricane.

2. The only tangential airflow component of "solid body rotation" type exists in the inner, core region, i.e.

$$u_\varphi = \omega_i r, \ \omega_i = const \ (0 < r \leq r_i). \tag{2.1}$$

According to the adiabatic approach and assumption 1 the vertical mass airflow being fed from the HBL, is conserved within the EW jet region, even if 3D airflow exists in this region. Also according to this model, only two wind speed components, tangential and radial ones, can exist in the external region $r > r_e(z)$. Although assumption 2 is well documented (e.g. see Refs. [7-12]) there are some unstable flows in the region 1, which seemingly maintain simplistic mean velocity distribution (2.1) in this region (e.g. see a comprehensive review in [13]).

The aerodynamic equations in EW region will be used in sequenced modeling in simplified form, where the density and vertical component of velocity are assumed to be radius independent (or averaged over the EW jet cross section), i.e.

$$u_z \approx u_*(z), \ \rho \approx \rho_*(z). \tag{2.2}$$

This simplification was introduced almost 60 years ago by Deppermann [7, 8]. Using additional empiricism, he was able to successfully describe the averaged behavior of typhoons. A possible reason for simplifications (2.2) is a high level of turbulent mixing of air in the EW jet.

The third simplification employed in the modeling is using vertical momentum balance equation averaged over EW cross section.

With the use of the above assumptions and simplifications the set of modeling aerodynamic equations in the EW region is presented as:

$$\rho_*\partial_r(ru_r)+r\partial_z(\rho_*u_*)=0, \quad u_r\partial_r M+u_*\partial_z M=0 \ ; \quad M=ru_\varphi+r^2\Omega. \tag{2.3}$$

$$\partial_r u_r^2/2+\rho_*^{-1}\partial_r p=\chi r \ ; \quad \chi=(\omega-\Omega)^2 \tag{2.4}$$

$$d/dz(\rho_*u_*\Delta s)=0 \quad \left[\Delta s(z)=r_e^2(z)-r_i^2\right] \tag{2.5}$$

$$2\frac{d}{dz}(\rho_*u_*^2\Delta s)=\frac{d}{dz}(p_a\Delta s-I_p)+g(\rho_a-\rho_*)\Delta s, \quad I_p=2\int_{r_i}^{r_e}rp\,dr. \tag{2.6}$$

Here $u_z\approx u_*(z)$, $u_r$ and $u_\varphi$ are respectively the vertical, radial and tangential wind speed components, $M$ is full angular momentum, $\omega$ and $\Omega$ are the angular velocities of relative hurricane rotation and local Earth rotation on $\beta$-plane, $p(r,z)$ is the local pressure, and $\rho_a(z)$ and $p_a(z)$ are ambient density and pressure described by barometric adiabatic distributions (1.6) from [1]. Approximations (2.2) were used in (2.3) – (2.6).

Equation (2.5) describing the vertical mass balance and the left-hand side of (2.6) are obtained from the first equation in (2.3) and stationary equation for the vertical momentum balance (1.9) [1] using procedure of averaging over EW cross section. In this procedure the natural kinematical conditions at the jet surfaces have been used:

$$u_r\big|_{r=r_i}=0; \quad u_*r_e{}'=u_r\big|_{r=r_e}. \tag{2.7}$$

The first condition in (2.7) means that the inner boundary of EW jet $r=r_i$ is non-penetrable for radial motions. Using the first model assumption along with the continuity equation (2.3) in the outer region $r>r_e(z)$, the second condition in (2.7) is interpreted as: the radial flow in the outer region is generated (or *induced*) by the slope of external boundary $r=r_e(z)$ of EW jet.

Formulas (2.5), (2.6) demonstrate the well known "jet" approach [14], which works surprisingly well for finding approximate solutions of complicated problems.

In order to avoid the Kelvin-Helmholtz instability, all the distributions of wind and pressure fields should be continuous at the boundaries of the three horizontal regions of hurricane upper layer. It is impossible, however, to avoid discontinuities of density and vertical velocity component in the model. Although one can use a procedure to smoothen these discontinuities, it seems that they properly reflect the main physical effect – turbulization of the air flow near those boundaries,

## 2.3. Distributions of velocity components and pressure

Approximations (2.1) and (2.2) with the use of conditions (2.7) allow easily obtaining solutions of equations (2.3) and (2.4) for the density, air flow and pressure distributions.

Model density distribution in the upper hurricane layer $z > h_b$ is:

$$\rho \approx \begin{cases} \rho_a(z), & (0 \leq r < r_i) \\ \rho_*(z) = \rho_a(z) T_a^0 / T_0, & (r_i < r < r_e) , \\ \rho_a(z), & (r > r_e) \end{cases} \quad \rho_a(z) = \rho_a^0 \psi^{2.5}, \quad \psi(z) = 1 - \frac{gz}{3.5 R T_a^0} \quad (2.8)$$

Formula (2.8) takes into account the barometric formulas (1.6) in [1], and also the fact that $T_0 > T_a^0$, i.e. that the average density in EW jet region is slightly warmer that outside the jet. In (2.8), $\rho_a^0$ and $T_a^0$ are the values of ambient density and Kelvin temperature at the surface of ocean, $T_0$ is the average temperature at the lover boundary, $z = h_b$ of the upper layer, and numerical parameters are introduced in the last two formulas in (2.8) using value of the adiabatic exponent $\gamma = 1.4$.

Model vertical velocity distribution in the upper hurricane layer $z > h_b$ is:

$$u_z \approx \begin{cases} 0, & (0 \leq r < r_i) \\ u_*(z), & (r_i < r < r_e) \\ 0, & (r > r_e) \end{cases} . \quad (2.9)$$

Model radial velocity distribution in the upper hurricane layer $z > h$ calculated from the continuity equation (2.3) is:

$$u_r(r,z) = \begin{cases} 0, & (0 \leq r \leq r_i) \\ u_{re} \frac{r^2 - r_i^2}{\Delta s} \cdot \frac{r_e}{r}, & (r_i \leq r \leq r_e) \\ u_{re} \cdot \frac{r_e}{r}, & (r \geq r_e) \end{cases} , \quad u_{re} = u_*(z) r_e{'}(z) . \quad (2.10)$$

Using the second equation in (2.3) to calculate the distribution of angular momentum $M(r,z)$ in EW region and taking into account formulas (2.9) and (2.10) in this region, yields the equation:

$$\frac{r'_e r_e}{r} \cdot \frac{r^2 - r_i^2}{\Delta s} \cdot \frac{\partial M}{\partial r} + \frac{\partial M}{\partial z} = 0.$$

The general solution of this equation has the form of the first integral (1.12) in [1]:

$$M = a_1 f(\xi) + a_2, \quad \xi(r,z) = (r^2 - r_i^2)/\Delta s.$$

Here $a_1$ and $a_2$ are the coordinate independent parameters, and $f(\xi)$ is an arbitrary function [1]. The linear choice of $f(\xi)$ allowing easy physical interpretation additionally has a minimal property proved in Appendix 1. This linear solution, which satisfies continuous boundary conditions, $M\big|_{r=r_i} = M_i$ and $M\big|_{r=r_e} = M_e$ is presented as:

$$M(r,z) = (M_e - M_i)(r^2 - r_i^2)/\Delta s(z) + M_i \quad \{r_i < r < r_e\}. \tag{2.11}$$

Here $M_i$ and $M_e$ are coordinate independent (constant) parameters of the angular momentum distribution, which may be formally presented as:

$$M_e = const, \quad M_i = \alpha M_e \quad (\alpha = const). \tag{2.12}$$

To estimate the value of parameter $\alpha$ we first note that at the bottom of the upper layer, the smooth distribution of tangential velocity $u_\varphi$ in the entire eye/EW region $\{0 \le r \le r_{e0}\}$ is possible only if $M_i = (r_i/r_{e0})^2 M_e$ ($\equiv M_i^*$). In any other case there must be a kink in continuous $u_\varphi$ distribution, which should be supported by a tangential friction force $\tau$. When $r \to r_i + 0$, this friction force is presented as $\tau_+ = k[(M_i^* - M_i)/r_i]^2$, where $k$ is a positive friction factor. On the other hand, the same tangential friction force acting from the eye region at the interface $r = r_i - 0$ is equal to $\tau_- = k(M_i/r_i)^2$. Equalizing $\tau_-$ and $\tau_+$, keeping in mind the relation $M_i^* = (r_i/r_{e0})^2 M_e$, and assuming that on both sides of interface $r = r_i$ the air rotates in one direction, yields: $\alpha = (r_i/r_{e0})^2/2$. Thus using second equation in (2.3) with account of formulas (2.1), (2.11)-(2.13) yields the following continuous space distribution of angular momentum in the upper hurricane layer, $z > h$:

$$M(r,z) \approx M_e \begin{cases} \alpha r^2/r_i^2, & (r_i \le r \le r_e) \\ \alpha + (1-\alpha)(r^2 - r_i^2)/\Delta s(z), & (r_i \le r \le r_e) \\ 1, & (r \ge r_e) \end{cases}; \quad \alpha = (r_i/r_{e0})^2/2. \tag{2.13}$$

Using (2.3) and (2.13) yields the space distribution of tangential velocity in the hurricane upper layer:

$$u_\varphi(r,z) \approx -\Omega r + M_e \begin{cases} \alpha r / r_i^2, & (r_i \leq r \leq r_e) \\ \alpha / r + (1-\alpha)(r - r_i^2/r)/\Delta s(z), & (r_i \leq r \leq r_e) \\ 1/r, & (r \geq r_e) \end{cases} \qquad (2.14)$$

Also due to (2.14) the space distribution of $\omega(r,z) = u_\varphi(r,z)/r$, the angular velocity relative to the local Earth rotation $\Omega$, is:

$$\omega(r,z) \approx -\Omega + M_e \begin{cases} \alpha / r_i^2, & (r_i \leq r \leq r_e) \\ \alpha / r^2 + (1-\alpha)(1 - r_i^2/r^2)/\Delta s(z), & (r_i \leq r \leq r_e) \\ 1/r^2, & (r \geq r_e) \end{cases} \qquad (2.15)$$

Approximation of radius independent density distribution in (2.2), and similar for the inner and outer regions, can be justified when using radial distributions (2.10) and (2.13), along with the adiabatic pressure-density relation in equation (2.4) for radial pressure distribution. E.g. in the outer region, the approximation $\rho \approx \rho_a(z)$ is satisfied with $\varepsilon$ - precision, where $\varepsilon$ ($\ll 1$) is presented as:

$$\varepsilon = \frac{\gamma - 1}{\gamma} \cdot \frac{(u_{\varphi e})^2 + u_{re}^2}{RT_a^0} \ .$$

Finally, the space distribution of pressure $p(r,z)$ can be calculated after integrating model equation (2.4) over the radius $r$ and using continuity conditions at the boundaries $r = r_i$ and $r_e(z)$ of the EW jet. The final awkward formulas are simplified by: (i) neglecting the terms with $\Omega$ (which is valid not far away from the EW jet), and (ii) neglecting the difference between $\rho_a(z)$ and $\rho_*(z)$. These simplified but still complicated formulas have the form:

$$p \approx \begin{cases} p = p_0 - \dfrac{\rho_a}{2}\dfrac{M_e^2}{s_e}\left(1 + \alpha^2 \dfrac{s_e(s_i - s)}{s_i^2} + s_e \dfrac{Z(s_i)}{(\Delta s)^2}\right), & (0 \leq s \leq s_i) \\[2mm] p_a - \dfrac{\rho_a}{2}\dfrac{M_e^2}{s_e}\left(1 + s_e \dfrac{Z(s)}{(\Delta s)^2}\right) - \dfrac{\rho_a u_{re}^2}{2}\dfrac{s_e}{s}\dfrac{(s - s_i)^2}{(\Delta s)^2}, & (s_i \leq s \leq s_e) \\[2mm] p_a - \dfrac{\rho_a}{2} \cdot \dfrac{s_e}{s}(M_e^2/s_e + u_{re}^2). & (s \geq s_e) \end{cases} \qquad (2.16)$$

Here $s = r^2$, $s_i = r_i^2$, $s_e = r_e^2(z)$, $\Delta s(z) = r_e^2 - r_i^2$; $p_a$ and $\rho_a$ are the ambient pressure and density described by the barometric adiabatic formulas, and the "hurricane depression" $p_0$ is the pressure at the axis of revolution of hurricane, and the function $Z(s)$ is:

$$Z(s) = (1-\alpha)^2(s_e - s) + (s_i - \alpha s_e)^2 \frac{s_e - s}{s s_e} - 2(1-\alpha)(s_i - \alpha s_e)\ln\frac{s_e}{s} \qquad (2.17)$$

Several conclusions can be made from the distributions (2.10) and (2.13)-(2.15) of kinematical variables.

1. Formula (2.10) obtained by Deppermann [7, 8] shows that $u_r$ and the slope $r_e'$ of EW jet profile, have the same sign because the vertical velocity $u_*$ directed upward is positive. In particular if $r_e' < 0$, the radial velocity $u_r$ is directed inward of EW jet squeezing EW jet. If $r_e' < 0$ the radial velocity is directed outwards, widening EW jet. In these two alternatives, the radial airflow, induced by the variation of EW external boundary is continuously extended inside of EW jet providing the Kelvin-Helmholtz stability, without contribution in the vertical flow of air. It means that these two types of radial airflow cause either radial squeezing or expanding deformation of the EW jet.

2. Derivation of (2.13) established that $M_e = const$, which means the conservation of value of angular momentum at $r \geq r_e(z)$ for any elevation $z \geq h_b$.

3. Formulas (2.14) and (2.15) show that with increasing radius, both the tangential and relative angular velocities in the outer region $r > r_e$ decay to zero values, achieved at a certain radius $r_a$, where

$$r_a = \sqrt{M_e/\Omega}. \qquad (2.18)$$

It means that $r_a$ is the external radius of hurricane.

4. Using (2.15), the angular velocity of EW jet rotation $\omega_e = \omega|_{r=r_e}$ is defined as:

$$\omega_e = M_e/r_e^2(z) - \Omega. \qquad (2.19)$$

This formula can explain the change of rotation direction of hurricane from cyclonic (counter clock wise) at $z = h_b$ to anti-cyclonic at $z \to h_a$, if $r_e(z)$ highly increases when approaching the limit adiabatic height $h_a \approx 3.5 RT_a^0/g$.

Additionally, formulas (2.16) and (2.17) show the radial pressure increase from the hurricane center towards periphery, where the minimum pressure $p_0$ shown in (2.17) represents the hurricane depression in the upper layer of hurricane.

## 2.4. Vertical evolution of EW jet structure

*Formulation*

Vertical distributions of the basic variables have not yet been established. So this Section derives coupled equations describing the vertical evolution of the external boundary $s_e(z)$ and vertical velocity $u_*(z)$ in EW jet, based on equations (2.5) and (2.6). Integral $I_p$ in this equation, calculated with the aid of formulas (2.16) and (2.17), is presented as:

$$I_p = p_a(z)\Delta s - \rho_a(z) M_e^2 [F_\alpha(q) + \beta^2 b(q)]/2. \qquad (2.20)$$

Here

$$F_\alpha(q) = \frac{3 - 5\alpha^2 + 2\alpha}{2} - 2\frac{(1-\alpha)^2}{q} + \frac{3(1-\alpha)^2 - 4\alpha(1-\alpha)q + \alpha^2 q^2}{q^2}\ln(q+1)$$

$$\beta = \frac{u_{re}\sqrt{s_e}}{M_e}, \quad b(q) = \frac{q(q-2) + \ln(q+1)}{2q^2}, \quad \alpha \approx \frac{1}{2(q_0+1)}, \quad q_0 = q\big|_{z=h_b} \qquad (2.21)$$

The expression for derivative $dF_\alpha/dq\big|_{\alpha=const} = F_\alpha{}'(q)$ has the form:

$$F_\alpha{}'(q) = 2\frac{(1-\alpha)^2}{q^2} + \frac{3(1-\alpha)^2 - 4\alpha(1-\alpha)q + \alpha^2 q^2}{q^2(1+q)} - 2\frac{(1-\alpha)(3 - 3\alpha - 2\alpha q)}{q^3}\ln(q+1). \quad (2.22)$$

As shown in Section (2.5), in realistic cases contribution of radial velocity to $I_p$ is negligible. Then substituting (2.20) with $\beta \to 0$ into (2.6) yields instead (2.8) the finite form of vertical mass balance (2.5),

$$\rho_a(z)q(z)u_*(z) = \rho_a(h)q_0 u_*(h). \qquad (2.23)$$

Additionally, the averaged vertical momentum balance in the EW has the form:

$$\frac{d}{dz}\left(\frac{u^2}{2}\right) = \frac{M_e^2}{2s_i q \rho_a(z)}\frac{d}{dz}[\rho_a(z)F_\alpha(q)] + g\left(\frac{T_0}{T_a^0} - 1\right). \qquad (2.24)$$

The last term in the right-hand side of (2.24) describing buoyancy effect, occurs only because of difference between the density $\rho_*(z)$ in the warm EW jet and ambient density $\rho_a(z)$, shown in (2.8). Neglecting the terms proportional to parameter $\beta$ in (2.21) means that the effect of radial velocity on the evolution of EW jet profile is negligible as compared to the effect of tangential velocity.

We now introduce the non-dimensional geometric and kinematic variables, pressure, and temperature, shown below by tildes:

$$\hat{z} = \frac{gz}{3.5RT_a^0}, \quad h = \hat{z}(h_b), \quad \hat{r} = \frac{r}{r_i}, \quad s = \hat{r}^2, \quad \hat{r}_a = \frac{r_a}{r_i}, \quad \hat{r}_e(z) = \sqrt{q(z)+1}$$

$$\hat{M} = \frac{M(r,z)}{M_e}, \quad \hat{\omega} = \frac{\hat{M}}{\hat{r}^2}, \quad \hat{\Omega} = \frac{r_i^2 \Omega}{M_e}, \quad \hat{u}_\phi = \frac{u_\phi r_i}{M_e}$$

$$\hat{u}_z = \frac{u_z(z)}{\sqrt{RT_a^0}}, \quad \hat{u}_{re} = \hat{u}_z r'_e(z), \quad \hat{u}_r = \frac{u_r(r,z)}{\sqrt{RT_a^0}}$$

$$\hat{p} = \frac{p(r,z)}{p_a^0}, \quad \hat{T}_a(\hat{z}) = \frac{T_a(z)}{T_a^0} = 1 - \hat{z}, \quad \hat{T}_0(\hat{z}) = \frac{T_0}{T_a^0}(1-\hat{z})$$

(2.25)

Here the adiabatic expressions have been used for non-dimensional temperature $\hat{T}$ and pressure $\hat{p}$. Additionally, two non-dimensional parameters $\mu$ (or $\mu_0$) and $\tau_0$ are introduced as follows:

$$\mu = \frac{M_e^2}{2s_i RT_0} = \mu_0 \cdot (q_0 + 1), \quad \mu_0 = \frac{M_e^2}{2s_{e0} RT_a^0} = \frac{(u_{\phi e}^0)^2}{2RT_a^0}, \quad \tau_0 = 3.5\left(\frac{T_0}{T_a^0} - 1\right) \quad (2.26)$$

Here parameter $\sqrt{2\mu_0}$ represents the non-dimensional maximal tangential speed at the bottom of EW jet. Using (2.25) and (2.26), equations (2.23) and (2.24) are presented in non-dimensional form as a coupled set:

$$\hat{u}q(1-\hat{z})^{2.5} = \hat{u}_0 q_0 (1-\hat{h})^{2.5}, \quad \frac{d}{d\hat{z}}\left(\frac{\hat{u}^2}{2}\right) = \frac{\mu}{q(1-\hat{z})^{2.5}} \frac{d}{d\hat{z}}\left((1-\hat{z})^{2.5} F_\alpha(q)\right) + \tau_0 \quad (\hat{z} \geq h). \quad (2.27)$$

The first equation in (2.27) expresses the conservation of mass flux in the EW jet, and the second the averaged vertical momentum balance. The first term in right-hand side of the second equation in (2.27) proportional to parameter $\mu$, describes the effect of jet rotation, and the second the buoyancy effect. Expressing $\hat{u}(\hat{z})$ from the first equation in (2.27) and substituting it into the second one, yields the boundary problem for vertical evolution of shape of EW jet, described by variable $q(\hat{z})$:

$$\frac{dq}{d\hat{z}} = \frac{q}{\psi} \cdot \frac{2.5u_0^2 q_0^2 \psi_h^5 + 2.5\mu q F_\alpha(q)\psi^5 - \tau_0 q^2 \psi^6}{u_0^2 q_0^2 \psi_h^5 + \mu q^2 F_\alpha{}'(q)\psi^5} \quad (\hat{z} \geq \hat{h}); \quad q(\hat{h}) = q_0 \tag{2.28}$$

$$\left(\psi = 1-\hat{z},\ \psi_h = 1-\hat{h},\ q_0 = q(\hat{h}),\ u_0 = \hat{u}(\hat{h}) > 0\right)$$

Hereafter the tildes are omitted and the value $\gamma = 1.4$ was used for adiabatic index. As soon as the solution of problem (2.28) is found, all the vertical/radial distributions of pressure and components of horizontal wind are easily calculated using the above analytical formulae. In particular, the vertical velocity $u(z)$ is calculated as:

$$\hat{u}(\hat{z}) = u_0 \frac{q_0}{q(\hat{z})} \cdot \left(\frac{\psi_h}{\psi(\hat{z})}\right)^{2.5}. \tag{2.29}$$

*Analysis*

To solve the problem (2.28) numerically one has to analyze the boundary value $q\,'(h)$ for $q\,'(z)$ in the limit $z \to h$. Hereafter we will use in this Section non-dimensional variables (2.25) omitting tildes. The magnitude $q\,'(h)$ is determined by the values of parameters $\mu_0$ and $\tau_0$, as well as by "initial" values of vertical velocity $u_0$ and geometrical parameter $q_0$, which can be interdependent. Equation (2.28) in the limit $z \to h$ becomes:

$$q\,'(h) \equiv \left(\frac{dq}{dz}\right)_{z=h} = -\frac{\tau_0 - [2.5u_0^2 + 2.5\mu_0(1+/q_0)\varphi_1(q_0)]/\psi_h}{u_0^2/q_0 + \mu_0(1+1/q_0)\varphi_2(q_0)} \quad \left(q_0 = \frac{s_e - s_i}{s_i}\right). \tag{2.30}$$

Functions $\varphi_1$ and $\varphi_2$ in (2.30) are defined as $\varphi_1(q_0) = F_\alpha(q)\big|_{q=q_0}$ and $\varphi_2(q_0) = F\,'_\alpha(q)\big|_{q=q_0}$. Their properties are described in Appendix 2.

Parameters $q_0, \mu_0, \tau_0$, and $u_0$ are considered in this paper as known. In fact, parameters $\mu_0, \tau_0$ and $u_0$ are determined by complicated processes in (quasi) stationary hurricane boundary layer, which will be analyzed in the next paper. Parameter $q_0$ can, however, be analyzed when considering non-stationary process of forming hurricane eye frontal surface at $r = r_i$. Using some realistic values of these parameters, it will be shown in the next Section that the values $u_0^2$ are negligible as compared to the values of other parameters in equation (2.30). Then simplified version of equation (2.30) is written as:

$$q'(h) \approx -(\tau_0 - \tau_1)/\tau_2. \qquad (2.31a)$$

$$\tau_1 = 2.5\mu_0(1+/q_0)\varphi_1(q_0)/\psi_h, \quad \tau_2 = \mu_0(1+1/q_0)\varphi_2(q_0). \qquad (2.31b)$$

The relation between $q'(h)$ and the slop of EW jet $r'_e(h)$ at the bottom of adiabatic layer is easy established as:

$$r'_e(h) = \kappa q'(h), \quad \kappa = \frac{g}{7RT_a^0\sqrt{q_0+1}}. \qquad (2.32)$$

Evidently, a stationary hurricane is stable only if $r'_e(h) < 0$, because only in this case the heat and moisture fluxes generated by induced radial flow in boundary layer, are centripetal. Thus due to (2.31a,b) and (2.32) the condition for hurricane stability is:

$$\tau_0 > \tau_1 \quad (q'(h) < 0). \qquad (2.33)$$

It means that the value $\tau_1$ in (3.7) is a critical value, which buoyancy should overcome to propagate upwards the stable EW jet against the action of cooling vertical adiabatic gradient. Interestingly enough, numerical simulations showed that when $\tau_1$ is nullified, the EW jet collapses with $q$ getting a negative value. I.e. in this case the eye region ceased to exist at a certain height. It means that the presence of critical term $\tau_1$ in (2.32) caused by adiabatic cooling gradient, also stabilizes the vertical EW jet.

If the values of $\tau_0$ ($> \tau_1$) and $u_0$ are known, one can also establish the value of maximal radial velocity at the bottom of adiabatic layer from the kinematical condition: $u_{re}(h) = u_0 r'_e(h)$. This relation, along with equation (2.30) or its simplified version (2.31), shows that the modulus of initial slop $|r'_e(h)|$ and modulus of initial radial velocity increases with the increase in the buoyancy parameter $\tau_0$. It means in turn that increasing buoyancy parameter increases the centripetal radial transport of temperature.

Inequality $q'(h) < 0$ means that $q'(z) < 0$ at least near the bottom of adiabatic layer, i.e. the hurricane EW jet is compressed there and the radial airflow is converging. In order to reveal what the hurricane structure might be at higher altitudes we now formally consider the asymptotic solution of equations (2.28), (2.29) at the limit $z \to 1$. Neglecting in equation (2.28) all the terms proportional to $(1-z)^5$ and $(1-z)^6$ reduces this equation to the following asymptotic form:

$$q`(z) \approx 2.5\psi_h^5 q/(1-z).$$

Then at $z \to 1$ asymptotic solutions of (2.28) and (2.29) have the form:

$$q = c_1\left(\frac{1-h}{1-z}\right)^{2.5\nu}, \quad u = c_2\left(\frac{1-h}{1-z}\right)^{2.5(1-\nu)}, \quad u_m = c_3\left(\frac{1-z}{1-h}\right)^{2.5\nu}, \quad v = (1-h)^5 \ (z \to 1). \quad (2.34)$$

Here $u_m$ is the mass velocity, and only one of three positive constants $c_k$ in (2.34) is arbitrary. Due to (2.34), the slop of vertical distribution $q(z)$ is positive, i.e. the hurricane EW jet near the limit adiabatic height is expanded, and the mass velocity goes to zero. It means that in the vicinity $z = 1$, vertical flow changes to the radial one. To obtain a complete solution in the whole region $0 < z < 1$ asymptotic solutions (2.34) should be matched with the solution of (2.38) when $z \to h$. Instead, we demonstrate in the last Section 2.6 of this paper some results of numerical simulations of EW jet profile $r_e(z)$.

Two physical conclusions can be made from the analyses of this Section.
1. The stability conditions (2.33) show that at the bottom of EW jet its external radius $r_e(z)$ decreases with height, inducing the converging radial flow from ambient air. On the contrary, close to the ultimate adiabatic height $h_a$ the external radius $r_e(z)$ of EW jet highly increases with height, injecting the radial flow from jet into the ambient air. It seems that this divergent radial flow is unstable.
2. The current model also predicts the change in the sign of angular velocity $\omega_e(z)$ of hurricane rotation depending on altitude. This is evident from formula (2.19) and the results of the above first conclusion. Indeed, according to first formula in (2.34) $r_e \to \infty$ at the adiabatic limit height, so in this limit $\omega_e = M_e / r_e^2(z) - \Omega < 0$.

Both the above model predictions correspond to the facts found in observations of hurricanes (e.g. see Refs.[9], [10]).

## II.5. Model parameters and numerical illustrations

*Parameters and estimations*

The common values of fundamental atmospheric parameters are [2, 7-10]:

$$R \approx 2.9 \cdot 10^2 \, J/(K^0 kg), \quad T_a^0 \approx 3 \cdot 10^2 \, K^0, \quad g \approx 9.81 m \cdot s^{-2}, \quad \Omega \approx (3-6) \times 10^{-6} s^{-1}. \tag{2.35}$$

Here $\Omega = \Omega_e \sin \beta$, and $\Omega_e \approx 1.16 \times 10^{-5} s$. The value $\Omega$ in (2.35) is estimated for the latitudes $\beta = 15^0 - 30^0 N$ common for Atlantic hurricanes, with the larger values of $\Omega$ being more suitable for the East Pacific typhoons.

Due to (2.35) the characteristic vertical adiabatic scale is:

$$l_z \approx RT_a^0 / g. \tag{2.36}$$

It is equal to $\approx 9,000 m$. The temperature at this altitude $T_{l_z} \approx 223^0 K = -50^0 C$, seems to be quite realistic. The dimensional values of temperature difference $T_0 - T_a^0 = (2-10)^0 C$ will be accepted for estimating the buoyancy parameter $\tau_0$ in (2.26) and (2.27).

The characteristic time $t_z$ of vertical restructuring in hurricanes is estimated as:

$$t_z = l_z / \sqrt{RT_a^0} = \sqrt{RT_a^0} / g. \tag{2.37}$$

Here $\sqrt{RT_a^0}$ is the characteristic scale of hurricane velocity. According to (2.35) $\sqrt{RT_a^0} \approx 295 m/s$ and $t_z = 30 s$, i.e. the vertical restructuring of hurricanes happens in a period of ~ 1min.

Hurricanes have a relatively low intensity of angular velocity, but a huge radial scales, with the external radius of EW jet at the bottom, $r_{e0} \approx 30 - 40 km$ and $r_a \approx 400 - 600 km$. A typical value of angular momentum $M_e$ in hurricanes is then estimated as $M_e = \Omega r_a^2 \approx 2 \times 10^6 \, m^2/\sec$. Thus at the hurricane jet radius $r_e = 4 \times 10^4 m$, the maximal rotational velocity is $u_{\varphi e} \equiv u_\varphi \big|_{r_e} \approx M_e / r_e \approx 50 m/\sec$.

Usually the observations of (quasi) stationary hurricanes provide the values for maximal horizontal velocity close to the rotational velocity $u_{\varphi e}$ and external hurricane EW radius $r_e$ at a certain altitude $z_{obs}$. If the value of $z_{obs}$ belongs to the interface of boundary and adiabatic layers, one can employ these values to calculate the angular momentum $M_e$ and external radius of hurricane $r_a$ using formula (2.18), where the effect of $\Omega$ on $M_e$ is neglected. Using the above values of $u_{\varphi e}$ and $\Omega$ makes also possible to estimate from (2.19) the value of critical radius $r_{ec}$ of EW jet where $\omega_e = 0$. Formulas (2.18) and (2.19) show that $r_{ec} = r_a$, where $r_a$ is the total radius of hurricane. However, because of instability of diverging radial flow generated at much lower heights, the change of direction of hurricane rotation can happen earlier.

The values of ratio $\xi = r_i / r_{e0}$ characterizing the EW radial thickness at the bottom of adiabatic layer are barely known. We assume that $\xi \approx 0.3 - 0.7$, with the most accepted medium value $\xi = 0.5$. Using the typical dimensional values $h_b = (0.75 - 1.8)km$ for the height of boundary layer, the range of variation for its non-dimensional value $\hat{h}$ is calculated as $\hat{h} = (2.4 - 5.62) \times 10^{-2}$, so $\psi_h = (0.968 - 0.940)$.

The typical dimensional values of vertical velocity $u_{0d} = (1-5)m/s$ result in the evaluation $u_0 \sim 10^{-2}$. This allows neglecting terms $u_0^2$ in (2.30) and using values of $u_0$ within the range $(0.34 - 1.7) \times 10^{-2}$.

The values of radial velocity $u_{re}^0$ at the hurricane bottom are typically in the range of $|u_{re}^0| = (3 - 25)m/s$, although the values as high as 30m/s have also been seldomly observed. According to the modeling of this paper, being in accord with literature observations, the greatest contribution of radial velocity in the lower part of adiabatic layer is at its bottom $(z = h_b)$, i.e. at the radius $r_{e0}$. In higher altitudes radial velocity quickly decays unless approaching to the adiabatic limit height. To estimate the contribution of radial velocity in the right-hand side of equation (2.28) we consider the realistic values of non-dimensional geometrical variables at $z = h_b$ as: $r_i / r_{e0} = 1/2$, $\alpha \approx 1/8$ and $q_0 = 3$. Then using typical maximal values for tangential and radial velocities $u_{\varphi e}^0 = 50 m/s$, $|u_{re}^0| = (5 - 25)$ $m/s$, yields $|\beta| \approx 0.02 - 0.5$. Calculating $F_\alpha(q)$ and $b(q)$ from (2.29) and (2.30) yields: $F_{1/8}(3) \approx 1.25$ and $b(3) \approx 0.244$. Thus the contribution of radial velocity in the right-hand side of equation (2.28), estimated as $\beta^2 b(3) / F_{1/8}(3)$ is equal to $\approx (0.02 - 5)\%$. Note that the highest error 5% of neglecting

contribution of radial velocity comes from the highest value of $\left|u_{re}^0\right| = 25 \, m/s$, which very seldomly observed in hurricanes.

*Choice of parameters for numerical illustrations*

It should be noted that values of basic physical parameters, $M, u_0$ and $\tau_0$ introduced below as independent, are interdependent in reality. Therefore the calculated results below cannot be considered more than qualitative illustrations.

**1.** The "standard" hurricane with the bottom geometry, $r_{e0} = 40 km$, $r_i = 20 km$ is considered in the first type of illustrations, where we accepted the values $u_{\varphi e}^0 = 50 m/s$ and $\Omega = 5 \times 10^{-6} 1/s$. In this case, $M_e = 2 \times 10^6 m^2/s$, and using (2.18) yields the hurricane external radius $r_a \approx 632 km$. We also use in illustrations a very low boundary layer height $h_b = 750 m$ and $T_a^0 = 300^0 K \, (27^0 C)$. Several examples of dimensional initial vertical velocities: $u_{0d} \approx 1$, 3 and $5 m/s$ will also be considered in calculations. The values of non-dimensional geometrical parameters used in illustrations are: $q_0 = 3$, $\alpha = 1/8$, $\varphi_1(q_0) \approx 1.25$, $\varphi_2(q_0) \approx 0.0329$, $\hat{h} \approx 0.0242$, $\psi_h \approx 0.976$, $\hat{r}_i \approx 0.6443$, $\kappa \approx 0.161$, and $\hat{r}_a \approx 31.6$. Non-dimensional dynamic parameters in equation (3.5) are: $\hat{\Omega} \approx 10^{-3}$, $\mu = 0.0575$ and $u_0 \approx 0.0102$. Using the values of parameters chosen above, the parameters $\tau_1$ and $\tau_2$ are calculated due to (2.31) as: $\tau_1 \approx 0.0598$ and $\tau_2 \approx 6.30 \times 10^{-4}$. The chose value of buoyancy parameter $\tau_0 = 0.07$ corresponds to the temperature $T_0 = 33^0 C$ (or $T_0 - T_a^0 \approx 6 \, ^0 C$) in the eye/EW region. Calculating with this value the right-hand side of full equation (2.30) gives the value of initial slope $q'(h) \approx -13.1$ or $r'_e(h) \approx -2.10$. The dimensional (or non-dimensional) value for initial radial velocity is: $u_{re}(h) \approx -10.5 m/s$ ($-0.0357$). The maximal dimensional value for horizontal wind speed is found as $V_{d\max}(h) = \sqrt{u_{\varphi\max}^2(h) + u_{re}^2(h)} \approx 51.0 m/s$, and respective non-dimensional value being equal to 0.510.

**2.** As the second type of illustrations, we calculated the data for the Deppermann typhoon model [7, 8]. In this model, the data with the hurricane of "standard" bottom geometry were considered as above, but with the height of boundary layer $h_b = 1.5 km$, meaning $\hat{h} \approx 0.0483$ and $\psi_h \approx 0.952$. Additionally, Deppermann used at this height the value for the full horizontal wind $V_d(h_d) = 55.6 m/s$ (~130mph), the dimensional value of radial velocity $u_{re}(h_d) = -30 m/s$ whose non-dimensional value is $\approx 0.102$, and the value of vertical velocity $3 m/s$ (the non-dimensional value $\approx 0.0102$). Bearing in mind that the Deppermann model uses the radial geometry of "standard" hurricane, the calculations give the values: $u_{\varphi \max}(h_d) = \sqrt{V_d^2(h_d) - u_{re}^2(h_d)} = 46.76 m/s$, $\mu = 0.05026$, $r'_e(h) = -10$, and $q'(h) \approx -62.07$. Using these data and making back calculations with fitting the above values of $q'(h)$ and $\tau_0 \approx 0.0914$, yields the value $T_0 = 34.8^0 C$ (or $T_0 - T_a^0 \approx 7.8\ ^0 C$).

*Numerical illustrations*

Figure 2.1 shows the schematics of the hurricane model. Figures 2.2 a,b illustrate the behavior of structural functions $\varphi_1(q_0)$ and $\varphi_2(q_0)$ described in Appendix 2. Other results of calculations of "standard" hurricane are presented in non-dimensional form in Figures 2.3–2.10, using the above fixed numerical values of parameters. Note that in these figures the vertical axis $z$ is counted off from the altitude corresponding to the beginning of adiabatic layer

Figure 2.3 demonstrates non-dimensional altitude dependence of outer boundary $r_e(z)$ of EW jet, obtained by numerical solution of boundary problem (2.28). Two values of dimensional initial vertical speed were used in calculations: $u_{0d} \approx 1$ and $3 m/s$. Their non-dimensional values are indicated in this Figure. It is seen that near the bottom the EW jet radius decreases with altitude. It achieves the minimal non-dimensional value $r_m \approx 1.87 (\approx 37.4 km)$ at non-dimensional altitude $z_m \approx 0.05$. This value $z_m$ corresponds the dimensional altitude $\approx 1.5 km$, counted off the height of boundary layer, or $\approx 3 km$ height counted from the oceanic surface. At $z > z_m$, the jet radius monotonically increases. At the higher altitude, $\approx 21.7 km$ corresponding to non-dimensional value $z = 0.7$, it is 1.8 times higher than the initial one, i.e. equal to 72 km.

Figure 2.4 shows the non-dimensional altitude dependences of vertical velocity $u(z)$ in the EW jet, obtained by solving problem (2.28) with the same values of $u_0$ as in Fig.2.3. The increase in $u(z)$ happens slowly enough: even at the higher altitude in our calculations, corresponding to $z = 0.7$, the dimensional value of vertical velocity with $u_{od} = 3 m/s$ is only $17 m/s$.

Figure 2.5 displays the vertical distribution of non-dimensional angular velocity $\omega_e(z)$ of EW jet according to formula (2.19). Here $r_e(z)$ was calculated with $u_0 \approx 0.017$ ($u_{0d} = 5 m/s$). It is seen that 11% increase in $\omega_e(z)$ achieved from initial value at $z = z_m$, changes to decrease at $z > z_m$, tending to acquire a negative value in the vicinity $z = 1$.

Figures (2.6ab) – (2.9ab) demonstrate radial distributions of non-dimensional angular momentum $M(r, z)$, tangential $u_\varphi(r, z)$ and radial $u_r(r, z)$ components of wind, and pressure $p(r, z)$, respectfully, at two non-dimensional altitudes: (a) $z = h$ and (b) $z = 0.3$ with given value of initial vertical velocity $u_0 \approx 0.017$ ($\approx 5 m/s$).

Figures 2.6a, b show monotonic increase of angular momentum in radial direction, but calculations also showed a non-monotonic evolution of the angular momentum with altitude near the inner boundary of EW, with slightly nonlinear increase in its value in the EW region toward periphery.

Figures 2.7a,b demonstrate the altitude change in position of tangential velocity peak. The value of this peak, initially changing non-monotonically, monotonously decreases at higher altitudes.

Figures 2.8a,b illustrate radial distributions of non-dimensional radial velocity This velocity is equal to zero in the eye region, its modulus increases in the EW region having the peak at the outer boundary $r_e$ of EW, and decreases as $\sim 1/r$ outside the EW region. Closed to the lower boundary of adiabatic layer, radial velocity is negative, and at altitudes higher than $z_m$ it changes the sign to be positive, with high enough maximum values.

When comparing the peak behavior of tangential and radial wind components presented in Figs. 2.7 and 2.8, it should be noted that due to (2.25) they have different scales. Of interests here is the ratio $\beta = |u_{re}|/u_{\varphi e}$, introduced in (2.21). This ratio steadily decreases with growing altitude. It means that the contribution of radial velocity in evolution of EW jet profile being small enough at $z = h_b$ decreases even more with growing altitude. This important result was confirmed by detailed numerical analysis.

Figures 2.9a,b demonstrate that all the radial pressure distributions have the similar features. They have monotonically increasing values of pressure from the minimal value $p_0$ of hurricane depression at $r = 0$, to the parabolic rise in the eye region followed by a steep increase in the EW region, and a gentle increase outside the EW region to the ambient values. Note that the ambient values $p_a^0(z)$ decrease themselves with altitude.

Figures 2.10a,b compare the Deppermann typhoon model [3] (red solid lines) and the present model (blue solid lines). Although Figure 2.10a indicates almost identical radial distributions of horizontal wind velocity for the present and Deppermann models, there are noticeable deviations between the two models in the surface pressure radial distributions, shown in Figure 2.10b. These pressure distributions almost coincide within the eye region, but the Deppermann's pressure distribution increases with radius steeper than that predicted by the present model, with the maximum deviation about 8mb at $r = r_e$. Note that Deppermann ignored by uncertain reason the contribution of radial velocity $u_r$ in the pressure. Therefore we also performed the calculation of pressure distribution without $u_r$ effect, presented in Figure 2.10b by the dotted blue line. This (dotted) distribution is much closer to the Depperman's one, with the maximum deviation about 2 mb.

## 2.6. Conclusions and discussions

The second paper of the series developed a simplified aerodynamic model which described and analyzed steady (or quasi-steady) 3D dynamic processes in the hurricane upper layer [1]. We remind that exact aerodynamic equations (1.7)-(1.10) in paper [1] describe hurricane as axisymmetric finite dynamic disturbances of the adiabatic barometric distributions of thermodynamic variables. Therefore it seemed reasonable to employ in the upper hurricane layer adiabatic approximation which ignores possible condensation processes in the layer. In accordance with observations, the structure of adiabatic layer was represented as three adjacent radial regions: eye, eye wall (EW) jet region, and outer region (Fig.2.1). The external boundary $r_e$ of EW jet depends on vertical coordinate $z$, whereas its internal boundary $r_i$ and external radius of hurricane $r_a$ were considered as constant.

To simplify exact aerodynamic equation the vertical airflows in the eye and outer regions of hurricane were assumed to be negligible. This modeling assumption along with adiabatic approximation views the rotating EW jet as dominant dynamic component in the adiabatic layer. It is overheated by the heat transfer in hurricane boundary layer and remains to be overheated in all altitudes as compared to the air in the outer zone, in spite of the adiabatic cooling happening in all three radial zones. The EW jet is ascending because of the buoyancy effect, and also exchanges the radial momentum with air in inner and outer regions of adiabatic layer.

Two additional simplifications are made to make the model analytically tractable. The first one is using radius-independent approximations (2.2) for vertical velocity $u_z$ and density $\rho$ in EW region, introduced long ago by Deppermann [7, 8]. A possible physical reason for approximations (2.2) is that the relatively low values $u_z$ in EW region are comparable with the values of turbulent pulsations, homogenizing radial distribution $u_z$. On the contrary, much higher values of other wind speed components make possible retaining their specific radial structure dictated by aerodynamics. Another essential simplification of the model is using radial average of equations describing the vertical momentum balance and continuity of EW jet, using the "jet approach" [14]. This makes possible to calculate the vertical distributions of both the vertical component of wind $u_z$ and the vertical evolution of the EW jet structure, i.e. $r_e(z)$. These simplifications result in the closed set of stationary axisymmetric equations (2.2)-(2.6) which constitute presented aerodynamic model.

Using (2.2) in the continuity equation yields radial distribution (2.10) for radial wind component $u_r$, which occurs because of inclination of the EW jet boundary $r_e(z)$.

Formula (2.10) coincides with that found in [7]. However unlike the further empiricism of paper [7], this paper establishes the expression (2.11) for angular momentum using the first integral (1.11) for steady and axisymmetric aerodynamic equations [1]. A fundamental feature of this solution is that the angular momentums $M_i$ at radius $r_i$ and $M_e$ at radius $r_e(z)$ of EW jet are constants. A relation between $M_i$ and $M_e$ was also found, and entire radial distribution of angular momentum is shown in (2.13). Using established velocity field and radial aerodynamic momentum balance, the model predicted the entire radial pressure distribution (2.16)/(2.17) in hurricane. All the above radial distributions are presented by analytical formulas. Inserting these distributions in the last equations (2.5) and (2.6) of the set yields the ODE problem (2.27), (2.28). They describe the vertical evolution of the external EW jet radius $r_e(z)$ and vertical wind speed component $u_z$. Although both the effects of jet rotation and radial component of wind speed affects the shape of EW jet and radial pressure distribution, the jet rotation effect highly dominates.

Analysis of equation (2.27) in Section 2.5 highlighted the dominant contribution of the temperature conditions and rotation of EW jet at its bottom on the vertical EW structure. It was shown that the EW jet is stable near the bottom of adiabatic layer, if the jet slope is negative. This happens only if the value of buoyancy parameter is high enough to overcome the action of cooling vertical adiabatic gradient. The negative slope of EW jet near the bottom induces the converging type of radial flow. Asymptotic solution (2.34) of equation (2.27) and (2.28), found for high altitudes near the adiabatic limit, show that in that region the air flow in EW jet is almost radial and diverging. This analysis, confirmed by our numerical simulations, also showed that the model correctly predicts two basic effects well known from the hurricane observations and not explained before. The first is the change in direction of rotation of angular velocity, from cyclonic in lower altitudes to ant-cyclonic in very high altitudes. The second is change of direction of radial velocity, from converging at low altitudes to diverging at higher altitudes.

Final Section of the paper estimated the basic parameters and use them in numerical illustrations of the basic results of paper. Here along with given bottom geometry of EW jet, the values of physical parameters were also introduced. These are the height $h_b$ of the HBL, the vertical velocity $u_{0d}$ at the bottom of EW jet, the constant angular momentum $M_e$, and a positive difference between the hot temperature $T_0$ within the eye/EW region and a critical one, $T_{0c}$ determined by the EW jet bottom geometry and rotation. These values were taken from the observational pool and (questionably) assumed to be independent. In fact they have to be found from analyses of processes in the HBL, which will be discussed in the next paper of the series. The numerical illustrations presented in Figures 2.3-2.9 seem to be realistic. The comparison between model calculations of distribution of horizontal wind speed and pressure, and corresponding plots obtained by Deppermann [7,9] is presented in Figures 10a,b and discussed at the end of previous Section. Since the Deppermann plots describe well the average behavior of many observed typhoons, this comparison seems encouraging.

**Appendix 1: Minimal property of function $f(\xi)$ describing angular momentum**

This Appendix shows that for any positive constants $a_1$ and $a_2$ and fixed values of $r_i$ and $r_e(z)$, the choice of linear function $f(\xi) = a_1\xi + a_2$ in expression of angular momentum $M(r,z)$ yields the minimal property for angular momentum functional within the class of smooth convex functions $f(\xi)$, under condition $f(1) = 1$. This functional problem is:

$$\forall z \geq 0, f''(\xi) > 0: \quad F(f) = 2\pi \int_{r_i}^{r_e(z)} [a_1 f(\xi) + a_2] r dr = \min; f(1) = 1, \ a_1, a_2 > 0. \quad (1A1)$$

The solution of this problem is evident if one introduces the Lagrange multiplier $v$ and considers the extremum of relative functional:

$$\tilde{F}(f) = 2\pi \int_{r_i}^{r_e(z)} \{a_1[f(\xi) - v] + a_2\} r dr = 2\pi \Delta s \left( a_1 \int_0^1 [f(\xi) - v] d\xi + a_2 \right). \quad (1A2)$$

The first variation of this functional gives the extremal condition:

$$\delta \tilde{F}(f) = 2\pi \Delta s a_1 \int_0^1 [f'(\xi) - v](\delta f(\xi)) d\xi = 0. \quad (1A3)$$

It means that $f(\xi) = \xi$ for any constants $a_1$ and $a_2$,. Calculation of the second variation of functional $\tilde{F}(f)$ shows that the extremum value $f(\xi) = \xi$ delivers minimum for this functional:

$$\delta^2 \tilde{F}(f) = 2\pi \Delta s a_1 \lim_{f(\xi) \to \xi} \int_0^1 f''(\xi)[\delta f(\xi)]^2 d\xi \geq 0. \qquad (1A4)$$

**Appendix 2: Behavior of structural functions $\varphi_1(q_0)$ and $\varphi_2(q_0)$**

Using definitions (2.21) and (2.22) for structural functions $F_\alpha(q)$ and $F_\alpha'(q)$ with $\alpha = 0.5/(q_0+1)$ one can present the functions $\varphi_1(q_0)$ and $\varphi_2(q_0)$ in (2.30) in the form:

$$\varphi_1(q_0) = \frac{1.5(0.5+q_0)\cdot(1.83333+q_0)}{(1+q_0)^2} - \frac{2(0.5+q_0)^2}{q_0(1+q_0)^2} + \frac{1.25(0.6+q_0)}{q_0^2(1+q_0)}\ln(1+q_0). \qquad (2A1)$$

$$\varphi_2(q_0) = \frac{2(0.5+q_0)^2 + 1.25(0.6+q_0)}{q_0^2(1+q_0)^2} - 2\frac{(0.5+q_0)(1.5+2q_0)]}{q_0^3(1+q_0)^2}\ln(1+q_0). \qquad (2A2)$$

We describe the behavior of these functions in the whole domain of their definition, $(0 < q_0 < \infty)$, bearing in mind a future analysis of hurricane degeneration.

Function $\varphi_1(q_0)$ behaves non-monotonically in the region where $q_0 \sim 1$. In the vicinity $q_0 \to 0$ asymptotic behavior of $\varphi_1(q_0)$ is singular and presented as $\varphi_1 \approx 0.25/q_0$. After achieving a maximum $\varphi_{1m} \approx 2.5$ at $q_{0m} \approx 1.5$, function $\varphi_1(q_0)$ then monotonically decreasing to the value $\varphi_1 = 1$ when $q_0 \to \infty$. Function $\varphi_1(q_0)(1+1/q_0)$ monotonously decreases, having in the vicinity $q_0 = 0$ the singular behavior $\sim 0.25/q_0^2$ and the limit value equal to unity when $q_0 \to \infty$. Plots of these functions are presented in Figure 2.2a.

Function $\varphi_2(q_0)$ also behaves non-monotonically and singularly at small values of $q_0 \ll 1$, and then monotonically decreases to limit $\varphi_2 = 0$ when $q_0 \to \infty$. Function $\varphi_2(q_0)(1+1/q_0)$ has similar behavior. These functions are presented in Figure 2.2b.

## Acknowledgment

Many thanks are given to my former Ph.D. Student Atanas Gagov for numerical simulations and graphics.

# Figure Captions

Fig.2.1. Schematics of hurricane

Fig.2.2: The graphs of structural functions, (a) $\varphi_1(q_0)$ and $\varphi_1(q_0)(1+1/q_0)$, and (b) $\varphi_2(q_0)$ and $\varphi_2(q_0)(1+1/q_0)$.

Fig. 2.3: Non-dimensional altitude dependence of outer boundary of EW jet $r_e(z)$. Parameters: $h = 0.02416$, $\mu = 0.0574$, $\tau_0 = 0.07$, $q_0 = 3$, $\Omega = 10^{-3}$, $u_0 = 0.01017$ (blue line) and $0.003627$ (red line).

Fig.2.4: Non-dimensional altitude dependence of vertical velocity $u(z)$ in the EW jet. Parameters are the same as in Fig.2.3

Fig.2.5: Vertical distribution of non-dimensional angular velocity $\omega_e(z)$ of EW jet; $u_0 = 0.01695$ ($u_{0d} = 5 m/s$).

Fig.2.6: Non-dimensional radial distributions of angular momentum $M(r,z)$ at two non-dimensional altitudes: (a) $z = h$, (b) $z = 0.3$.

Fig.2.7: Non-dimensional radial distributions of tangential velocity $u_\varphi(r,z)$ at two non-dimensional altitudes: (a) $z = h$, (b) $z = 0.3$.

Fig.2.8: Non-dimensional radial distributions of radial velocity $u_r(r,z)$ at two non-dimensional altitudes: (a) $z = h$, (b) $z = 0.3$.

Fig. 2.9: Non-dimensional radial distributions of pressure $p(r,z)$ at four non-dimensional altitudes: (a) $z = h$, (b) $z = 0.3$.

Fig.2.10: Comparison of calculations of present and Deppermann models. (a) radial distribution of horizontal wind speed $V = \sqrt{u_\varphi^2 + u_r^2}$ (in *m/s*) at the bottom of adiabatic layer: red solid line - Deppermann model, blue solid line – present model, (b) radial distribution of surface pressure: red solid line - Deppermann model, blue solid line – present model, blue dotted line – calculations of present model, ignoring the radial velocity contribution.

**FIGURES**

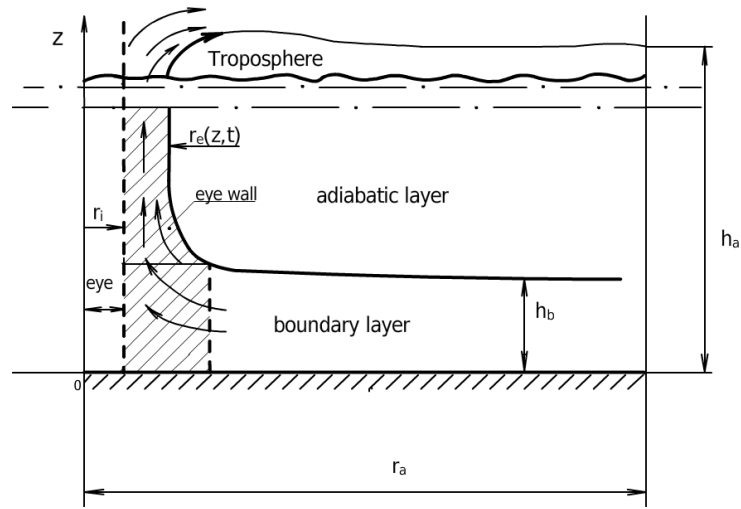

Fig.2.1

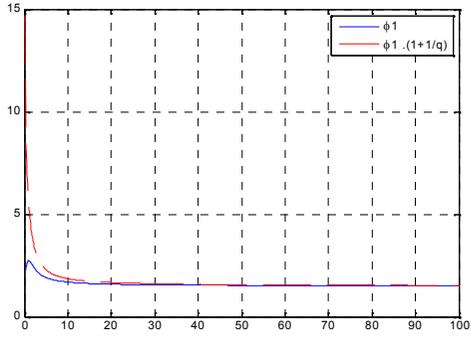
Fig. 2.2 a

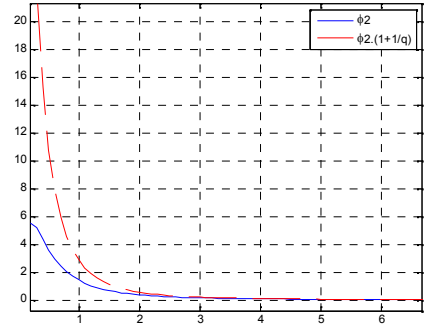
Fig.2.2 b

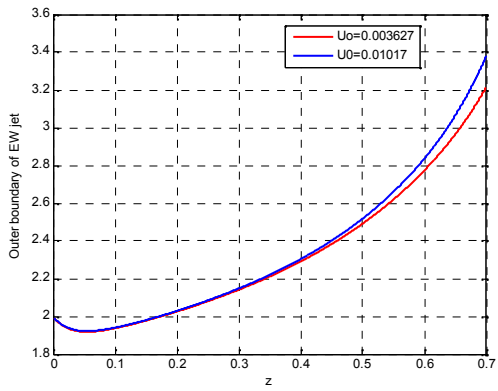
Fig. 2.3

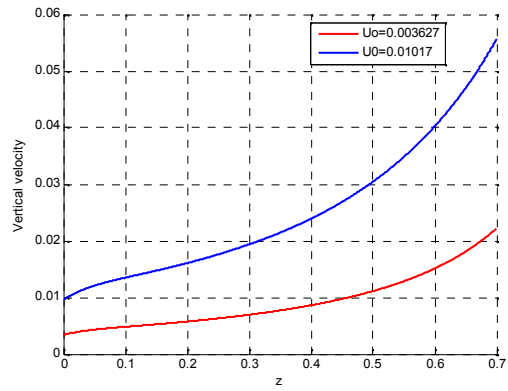
Fig.2.4

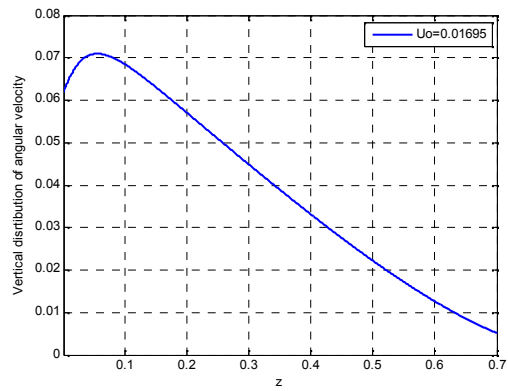
Fig. 2.5

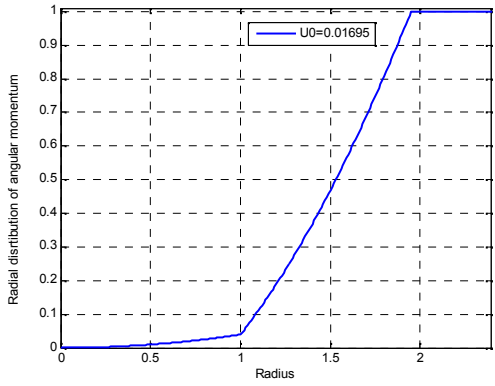
Fig.2.6 a

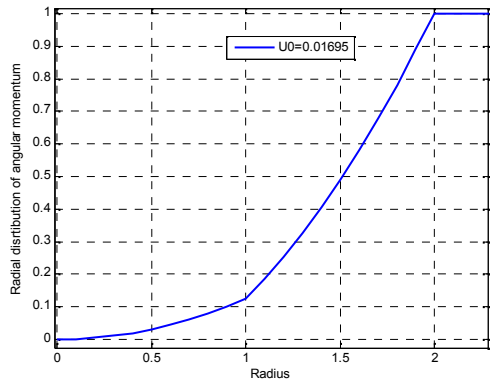
Fig.2.6 b

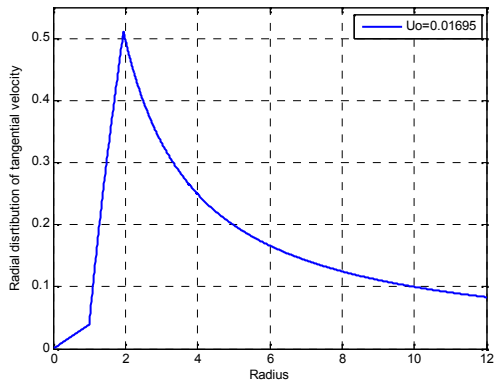
Fig.2.7 a

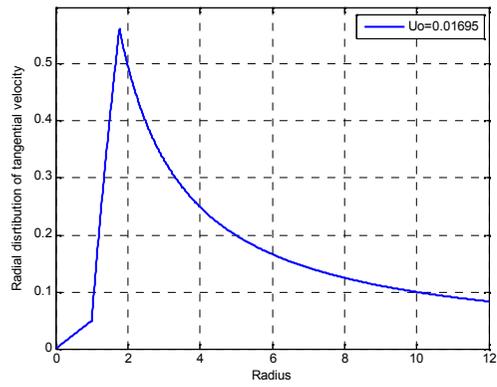
Fig.2.7 b

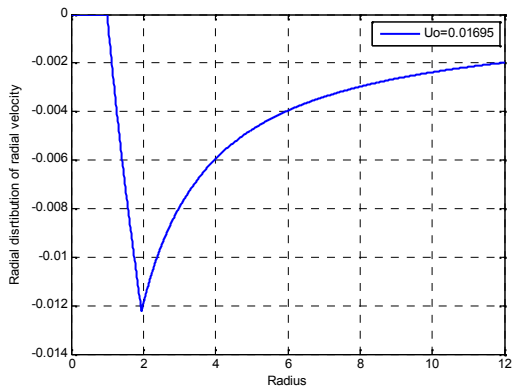
Fig.2.8 a

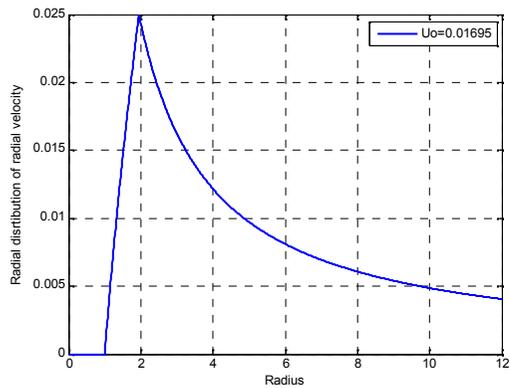
Fig.2.8 b

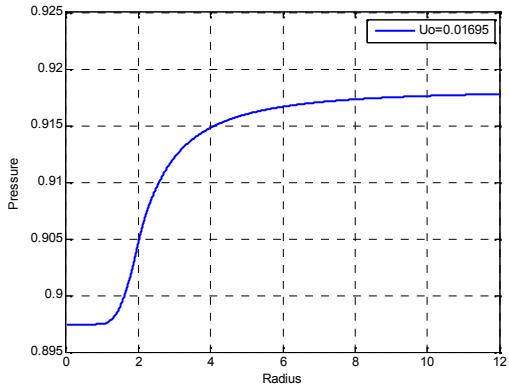
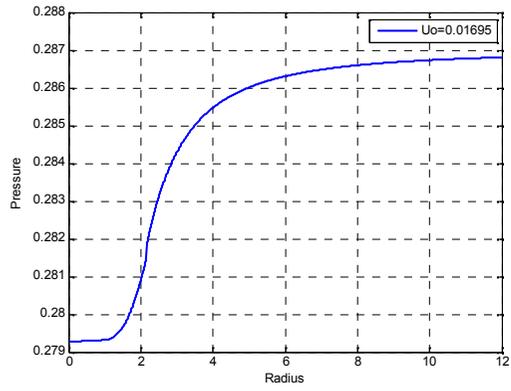

Fig.2.9 a

Fig.2.9b

## Deppermann Model

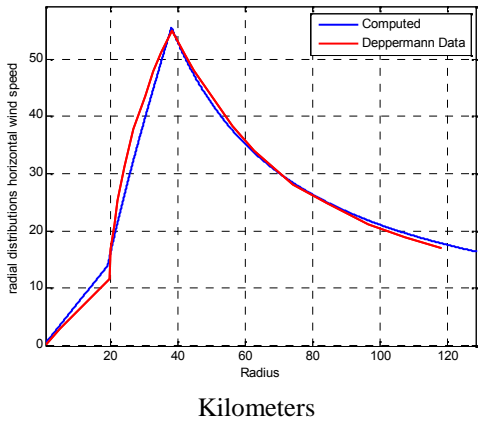
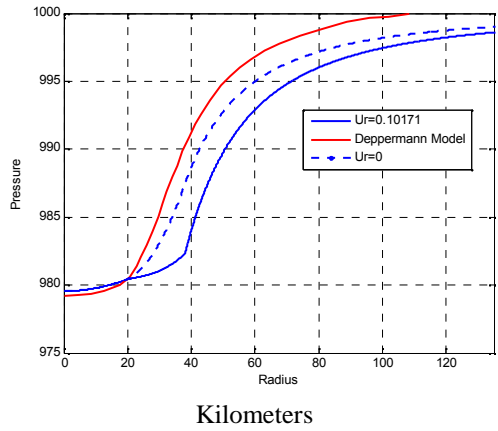

Fig.2.10 a

Fig.2.10 b